\documentclass[11pt]{article}
\usepackage{blindtext}
\usepackage{graphicx}
\usepackage{amsmath}
\usepackage{amssymb}
\usepackage{amsthm}
\usepackage{xkeyval}
\title{Quantum Analysis of Subharmonic Generation Via First-Order Hamiltonian}
\author{Fesseha Kassahun\footnote{Email address: fesseha.kassahun@aau.edu.et} \\
                Department of Physics,
                Addis Ababa University\\
                P. O. Box 33761, Addis Ababa, Ethiopia}

\begin{document}
\maketitle
\begin{abstract}
We have established that the mean photon number obtained employing the conventional Hamiltonian for the twin light beams with the same frequency, produced by  a subharmonic generator, is half that of the twin light beams with different frequencies. In view of this, we have found  it to be quite appropriate to describe the process of subharmonic generation by the same Hamiltonian regardless of whether the twin light beams have the same or different frequencies.  Applying this Hamiltonian, we have calculated the photon statistics and quadrature squeezing for the signal-signal and the signal-idler modes. Obviously, this time the mean photon number of the signal-signal modes is exactly the same as that of the signal-idler modes. In addition, we have found that the  maximum global quadrature squeezing of the signal-signal or the signal-idler modes is 50$\%$ below the vacuum-state level and the maximum local quadrature squeezing of the signal-signal modes is 74.9$\%$ below the same level.

\end{abstract}
\hspace*{9.5mm}Keywords: Hamiltonian, Photon statistics, Global and local quadrature squeezing
\vspace*{5mm}
\section{Introduction}
In subharmonic generation a pump photon of frequency $\omega$ is down converted into a pair of highly correlated  photons of frequencies $\omega_{1}$ and $\omega_{2}$, with $\omega_{1}$ and $\omega_{2}$ being the same or different.
The statistical and squeezing properties of the twin light beams  with the same or different frequencies (each light beam consisting of one photon from each pair)  have been investigated by several authors [1-4]. It is found that the twin light beams are in a squeezed state, with the maximum quadrature squeezing being 50$\%$ below the vacuum-state level. We recall that the twin light beams with the same frequency are represented in the conventional Hamiltonian by $\hat{a}^{2}$ and $\hat{a}^{\dag 2}$. For a given pump mode, we anticipate the mean  photon  number of the twin light beams with the same or different frequencies to be the same. However, the mean photon number of the twin light beams with the same frequency obtained, in the Appendix using the conventional Hamiltonian, is found to be half that of the twin light beams with different frequencies, as determined in section 4. We maintain the standpoint that the number operator $\hat{a}^{\dag}\hat{a}$ holds only for a light mode represented in the pertinent Hamiltonian by first-order annihilation and creation operators. The unexpected result we have obtained for the mean photon number of the twin light beams with the same frequency must then be due to representation of these twin light beams by second-order annihilation and creation operators.

We may envisage subharmonic generation as a process in which a three-level atom absorbs a pump photon and makes a transition from the bottom to the top level. Then it emits a photon and decays to the intermediate level. Finally, the atom emits another photon and decays to the bottom level. We certainly expect several three-level atoms to be involved in this process. The photons emitted from the top level constitute one light mode and those emitted from the intermediate level form another one. The two light modes may have the same or different frequencies. It therefore appears to be quite appropriate to describe the process of subharmonic generation leading to the creation of twin light modes with the same or different frequencies by the Hamiltonian [5]
\begin{equation}\label{61}
\hat H=i\mu(\hat b^{\dag}-\hat b)
+i\lambda(\hat b^{\dag}\hat{a}_{1}\hat{a}_{2}-\hat b\hat{a}^{\dag}_{1}\hat{a}^{\dag}_{2}),
\end{equation}
where $\hat{a}_{1}$ and  $\hat{a}_{2}$ are the annihilation operators for the light modes emitted from the top and intermediate levels, $\hat{b}$ is the annihilation operator for the pump mode, $\lambda$ is the coupling constant, and $\mu$ is proportional to the amplitude of the coherent light deriving the pump mode. We may refer to a Hamiltonian of the form described by (\ref{61}) as first-order Hamiltonian.
We assume that the operators $\hat{a}_{1}$ and $\hat{a}_{2}$ commute and satisfy the commutation relation
\begin{equation}\label{62}
[\hat{a}_{1},\hat{a}_{1}^{\dag}]=[\hat{a}_{2},\hat{a}_{2}^{\dag}]=1.
\end{equation}

 We next seek to calculate, applying the Hamiltonian given by Eq. (\ref{61}), the mean and variance of the photon number, the global  quadrature squeezing, and the photon number distribution for the twin light modes with the same or different frequencies. We also wish to obtain the local quadrature squeezing for the twin light modes with the same frequency.

\section{Operator dynamics}
\begin{picture}(200,80)(-235,45)
\put (-125,50){\line(0,1){60}}
\put (-124.5,50){\line(0,1){60}}
\put (50,50){\line(0,1){60}}
\put (-55,65){\framebox(90,30){nonlinear crystal}}
\put (50,85){\vector(1,0){80}}
\put(55,90){light mode $a_{1}$}
\put (50,75){\vector(1,0){80}}
\put(55,63){light mode $a_{2}$}
\put (-125,80){\vector(1,0){68.8}}
\put(-120,83){pump mode}
\put (-150,80){\vector(1,0){26}}
\put(-140,83){$\mu$}
\end{picture}

\begin{center}
{\footnotesize{\bf Fig.~6.1}~~~~~A subharmonic generator.}
\end{center}
\vspace*{3mm}
We consider here the case in which the twin light modes and the pump mode are in a cavity coupled to a vacuum reservoir via a single-port mirror.
It is not hard to realize that
\begin{equation}\label{63}
{d\over dt}\langle\hat{a}_{1}\rangle=-iTr\big(\rho[\hat{a}_{1},\hat{H}]\big)+{\kappa_{1}\over 2}Tr\big((2\hat{a}_{1}\hat\rho\hat{a}_{1}^{\dagger}-\hat{a}_{1}^{\dagger}
\hat{a}_{1}\hat\rho-\hat\rho\hat{a}_{1}^{\dagger}\hat{a}_{1})\hat{a}_{1}\big),
\end{equation}
\begin{equation}\label{64}\overleftarrow{}
{d\over dt}\langle\hat{a}_{2}\rangle=-iTr\big(\rho[\hat{a}_{2},\hat{H}]\big)+{\kappa_{2}\over 2}Tr\big((2\hat{a}_{2}\hat\rho\hat{a}_{2}^{\dagger}-\hat{a}_{2}^{\dagger}
\hat{a}_{2}\hat\rho-\hat\rho\hat{a}_{2}^{\dagger}\hat{a}_{2})\hat{a}_{2}\big),
\end{equation}
\begin{equation}\label{65}{d\over dt}\langle\hat{a}_{1}^{\dag}\hat{a}_{1}\rangle=-iTr\big(\rho[\hat{a}_{1}^{\dag}\hat{a}_{1},\hat{H}]\big)+{\kappa_{1}\over 2}Tr\big((2\hat{a}_{1}\hat\rho\hat{a}_{1}^{\dagger}-\hat{a}_{1}^{\dagger}
\hat{a}_{1}\hat\rho-\hat\rho\hat{a}_{1}^{\dagger}\hat{a}_{1})\hat{a}_{1}^{\dag}\hat{a}_{1}\big),
\end{equation}
\begin{equation}\label{66}{d\over dt}\langle\hat{a}_{2}^{\dag}\hat{a}_{2}\rangle=-iTr\big(\rho[\hat{a}_{2}^{\dag}\hat{a}_{2},\hat{H}]\big)+{\kappa_{2}\over 2}Tr\big((2\hat{a}_{2}\hat\rho\hat{a}_{2}^{\dagger}-\hat{a}_{2}^{\dagger}
\hat{a}_{2}\hat\rho-\hat\rho\hat{a}_{2}^{\dagger}\hat{a}_{2})\hat{a}_{2}^{\dag}\hat{a}_{2}\big),
\end{equation}
\begin{eqnarray}\label{67}
{d\over dt}\langle \hat{a}_{1}\hat{a}_{2}\rangle\hspace*{-3mm}&=\hspace*{-3mm}&-iTr\big(\rho[\hat{a}_{1}\hat{a}_{2},\hat{H}]\big)
+{\kappa_{1}\over 2}Tr\bigg(\big(2\hat{a}_{1}\hat\rho\hat{a}_{1}^{\dagger}-\hat{a}_{1}^{\dagger}
\hat{a}_{1}\hat\rho-\hat\rho\hat{a}_{1}^{\dagger}\hat{a}_{1}\big)\hat{a}_{1}\hat{a}_{2}\bigg)\nonumber\\&&
+{\kappa_{2}\over 2}Tr\bigg(\big(2\hat{a}_{2}\hat\rho\hat{a}_{2}^{\dagger}-\hat{a}_{2}^{\dagger}
\hat{a}_{2}\hat\rho-\hat\rho\hat{a}_{2}^{\dagger}\hat{a}_{2}\big)\hat{a}_{1}\hat{a}_{2}\bigg),
\end{eqnarray}
in which $\kappa_{1}$ and $\kappa_{2}$ are the cavity damping constants for light modes ${a}_{1}$ and ${a}_{2}$, respectively.
Now in view of Eq. (\ref{61}) and the fact that
\begin{equation}\label{68}
{\kappa_{1}\over 2}Tr\big((2\hat{a}_{1}\hat\rho\hat{a}_{1}^{\dagger}-\hat{a}_{1}^{\dagger}
\hat{a}_{1}\hat\rho-\hat\rho\hat{a}_{1}^{\dagger}\hat{a}_{1})\hat{a}_{1}\big)=-{\kappa_{1}\over 2}\langle\hat{a}_{1}\rangle,
\end{equation}
\begin{equation}\label{69}
{\kappa_{2}\over 2}Tr\big((2\hat{a}_{2}\hat\rho\hat{a}_{2}^{\dagger}-\hat{a}_{2}^{\dagger}
\hat{a}_{2}\hat\rho-\hat\rho\hat{a}_{2}^{\dagger}\hat{a}_{2})\hat{a}_{2}\big)=-{\kappa_{2}\over 2}\langle\hat{a}_{2}\rangle,
\end{equation}
\begin{equation}\label{610}
{\kappa_{1}\over 2}Tr\big((2\hat{a}_{1}\hat\rho\hat{a}_{1}^{\dagger}-\hat{a}_{1}^{\dagger}
\hat{a}_{1}\hat\rho-\hat\rho\hat{a}_{1}^{\dagger}\hat{a}_{1})\hat{a}_{1}^{\dag}\hat{a}_{1}\big)=-\kappa_{1}\langle\hat{a}_{1}^{\dag}\hat{a}_{1}\rangle,
\end{equation}
\begin{equation}\label{611}
{\kappa_{2}\over 2}Tr\big((2\hat{a}_{2}\hat\rho\hat{a}_{2}^{\dagger}-\hat{a}_{2}^{\dagger}
\hat{a}_{2}\hat\rho-\hat\rho\hat{a}_{2}^{\dagger}\hat{a}_{2})\hat{a}_{2}^{\dag}\hat{a}_{2}\big)=-\kappa_{2}\langle\hat{a}_{2}^{\dag}\hat{a}_{2}\rangle,
\end{equation}
\begin{eqnarray}\label{612}
&{\kappa_{1}\over 2}Tr\bigg(\big(2\hat{a}_{1}\hat\rho\hat{a}_{1}^{\dagger}-\hat{a}_{1}^{\dagger}
\hat{a}_{1}\hat\rho-\hat\rho\hat{a}_{1}^{\dagger}\hat{a}_{1}\big)\hat{a}_{1}\hat{a}_{2}\bigg)\nonumber\\
&+{\kappa_{2}\over 2}Tr\bigg(\big(2\hat{a}_{2}\hat\rho\hat{a}_{2}^{\dagger}-\hat{a}_{2}^{\dagger}
\hat{a}_{2}\hat\rho-\hat\rho\hat{a}_{2}^{\dagger}\hat{a}_{2}\big)\hat{a}_{1}\hat{a}_{2}\bigg)\nonumber\\
&=-{1\over 2}(\kappa_{1}+\kappa_{2})\langle\hat{a}_{1}\hat{a}_{2}\rangle,
\end{eqnarray}
we readily get
\begin{equation}\label{613}
{d\over dt}\langle\hat{a}_{1}\rangle=-{1\over 2}\kappa_{1}\langle\hat{a}_{1}\rangle
-\lambda\langle\hat{b}\hat{a}_{2}^{\dagger}\rangle,
\end{equation}
\begin{equation}\label{614}{d\over dt}\langle\hat{a}_{2}\rangle=-{1\over 2}\kappa_{2}\langle\hat{a}_{2}\rangle
-\lambda\langle\hat{b}\hat{a}_{1}^{\dagger}\rangle,
\end{equation}
\begin{equation}\label{615}{d\over dt}\langle\hat{a}_{1}^{\dag}\hat{a}_{1}\rangle=-\kappa_{1}\langle\hat{a}_{1}^{\dag}\hat{a}_{1}\rangle
-\lambda\langle\hat{b}^{\dag}\hat{a}_{1}\hat{a}_{2}\rangle-\lambda\langle\hat{b}\hat{a}_{1}^{\dag}\hat{a}_{2}^{\dag}\rangle,
\end{equation}
\begin{equation}\label{616}{d\over dt}\langle\hat{a}_{2}^{\dag}\hat{a}_{2}\rangle=-\kappa_{2}\langle\hat{a}_{2}^{\dag}\hat{a}_{2}\rangle
-\lambda\langle\hat{b}^{\dag}\hat{a}_{1}\hat{a}_{2}\rangle-\lambda\langle\hat{b}\hat{a}_{1}^{\dag}\hat{a}_{2}^{\dag}\rangle,
\end{equation}
\begin{equation}\label{617}
{d\over dt}\langle \hat{a}_{1}\hat{a}_{2}\rangle=-{1\over 2}(\kappa_{1}+\kappa_{2})\langle\hat{a}_{1}\hat{a}_{2}\rangle
-\lambda\langle\hat{b}\hat{a}_{1}^{\dag}\hat{a}_{1}\rangle-\lambda\langle\hat{b}
\hat{a}_{2}^{\dag}\hat{a}_{2}\rangle-\lambda\langle\hat{b}\rangle.
\end{equation}
On taking $\kappa_{1}=\kappa_{2}=\kappa$, the steady-state solutions of the above equations are found to be
\begin{equation}\label{618}
\langle\hat{a}_{1}\rangle=
-{2\lambda\over\kappa}\langle\hat{b}\hat{a}_{2}^{\dagger}\rangle,
\end{equation}
\begin{equation}\label{619}
\langle\hat{a}_{2}\rangle=
-{2\lambda\over\kappa}\langle\hat{b}\hat{a}_{1}^{\dagger}\rangle,
\end{equation}
\begin{equation}\label{620}
\langle\hat{a}_{1}^{\dag}\hat{a}_{1}\rangle=
-{\lambda\over\kappa}\langle\hat{b}^{\dag}\hat{a}_{1}\hat{a}_{2}\rangle
-{\lambda\over\kappa}\langle\hat{b}\hat{a}_{1}^{\dag}\hat{a}_{2}^{\dag}\rangle,
\end{equation}
\begin{equation}\label{621}
\langle\hat{a}_{2}^{\dag}\hat{a}_{2}\rangle=
-{\lambda\over\kappa}\langle\hat{b}^{\dag}\hat{a}_{1}\hat{a}_{2}\rangle
-{\lambda\over\kappa}\langle\hat{b}\hat{a}_{1}^{\dag}\hat{a}_{2}^{\dag}\rangle,
\end{equation}
\begin{equation}\label{622}
\langle\hat{a}_{1}\hat{a}_{2}\rangle=
-{\lambda\over\kappa}\langle\hat{b}\hat{a}_{1}^{\dag}\hat{a}_{1}\rangle-{\lambda\over\kappa}\langle\hat{b}
\hat{a}_{2}^{\dag}\hat{a}_{2}\rangle-{\lambda\over\kappa}\langle\hat{b}\rangle.
\end{equation}

Upon dropping the noise operator and in the absence of subharmonic generation $(\lambda=0)$, one can write the quantum Langevin equation for the operator $\hat{b}$ as
\begin{equation}\label{623}
{d\hat{b}\over dt}=-{1\over 2}\kappa\hat{b}+\mu,
\end{equation}
where $\kappa$ is the cavity damping constant.
The steady-state solution of this equation is
\begin{equation}\label{624}
\hat{b}={2\mu\over\kappa}.
\end{equation}
Now on introducing (\ref{624}) into Eqs. (\ref{618})-(\ref{622}), we arrive at
\begin{equation}\label{625}
\langle\hat{a}_{1}\rangle=
-{2\varepsilon\over\kappa}\langle\hat{a}_{2}^{\dagger}\rangle,
\end{equation}
\begin{equation}\label{626}
\langle\hat{a}_{2}\rangle=
-{2\varepsilon\over\kappa}\langle\hat{a}_{1}^{\dagger}\rangle,
\end{equation}
\begin{equation}\label{627}
\langle\hat{a}_{1}^{\dag}\hat{a}_{1}\rangle=
-{\varepsilon\over\kappa}\langle\hat{a}_{1}\hat{a}_{2}\rangle
-{\varepsilon\over\kappa}\langle\hat{a}_{1}^{\dag}\hat{a}_{2}^{\dag}\rangle,
\end{equation}
\begin{equation}\label{628}
\langle\hat{a}_{2}^{\dag}\hat{a}_{2}\rangle=
-{\varepsilon\over\kappa}\langle\hat{a}_{1}\hat{a}_{2}\rangle
-{\varepsilon\over\kappa}\langle\hat{a}_{1}^{\dag}\hat{a}_{2}^{\dag}\rangle,
\end{equation}
\begin{equation}\label{629}
\langle\hat{a}_{1}\hat{a}_{2}\rangle=
-{\varepsilon\over\kappa}\langle\hat{a}_{1}^{\dag}\hat{a}_{1}\rangle-{\varepsilon\over\kappa}\langle
\hat{a}_{2}^{\dag}\hat{a}_{2}\rangle-{\varepsilon\over\kappa},
\end{equation}
in which $\varepsilon$ is defined by
\begin{equation}\label{630}
\varepsilon={2\mu\lambda\over\kappa}.
\end{equation}

Applying Eqs. (\ref{625}) and (\ref{626}), we easily find
\begin{equation}\label{631}
\langle\hat{a}_{1}\rangle=\langle\hat{a}_{2}\rangle=0.
\end{equation}
Moreover, using Eqs. (\ref{627}), (\ref{628}), and (\ref{629}), one gets
\begin{equation}\label{632}
\langle\hat{a}_{1}^{\dag}\hat{a}_{1}\rangle={2\varepsilon^{2}\over{\kappa^{2}
-4\varepsilon^{2}}},
\end{equation}
\begin{equation}\label{633}
\langle\hat{a}_{2}^{\dag}\hat{a}_{2}\rangle=\langle\hat{a}_{1}^{\dag}\hat{a}_{1}\rangle,
\end{equation}
and
\begin{equation}\label{634}
\langle \hat{a}_{1}\hat{a}_{2}\rangle=-{\kappa\varepsilon\over{\kappa^{2}-4\varepsilon^{2}}}.
\end{equation}
It can also be readily verified that
\begin{equation}\label{635}
\langle\hat{a}_{1}^{2}\rangle=\langle\hat{a}_{2}^{2}\rangle=\langle\hat{a}_{1}^{\dag}\hat{a}_{2}\rangle=0.
\end{equation}
We take
\begin{equation}\label{636}
\hat{a}=\hat{a}_{1}+\hat{a}_{2}
\end{equation}
to be the annihilation operator for the superposition of light modes ${a}_{1}$ and ${a}_{2}$, produced by the subharmonic generator. One can easily check that
\begin{equation}\label{637}
[\hat{a},\hat{a}^{\dag}]=2.
\end{equation}
We realize that the superposition of the two light modes, with the same or different frequencies, constitutes a two-mode light. We wish to call the superposed light modes with the same frequency the signal-signal modes and the superposed light modes with differen frequencies the signal-idler modes. It also proves to be convenient to refer to the subharmonic generator which produces the signal-signal modes as the degenerate subharmonic generator and the one which produces the signal-idler modes as the nondegenerate subharmonic generator. Finally, we would like to mention that the results described by Eqs. (\ref{631})-(\ref{635}) are valid for the signal-signal or the signal-idler modes.

\section{The Q function}

We next seek to obtain the Q function for the two-mode subharmonic light (the signal-signal or the signal-idler modes). This Q function is expressible as
\begin{equation}\label{638}
Q(\alpha_{1},\alpha_{2},t)={1\over\pi^{4}}\int d^{2}zd^{2}\eta\phi_{a}(z,\eta,t)\exp(z^{*}\alpha_{1}-z\alpha_{1}^{*}+\eta^{*}\alpha_{2}-\eta\alpha_{2}^{*}),
\end{equation}
in which the antinormally-ordered characteristic function $\phi_{a}(z,\eta,t)$ is defined in the Heisenberg picture by
\begin{equation}\label{639}
\phi_{a}(z,\eta,t)=Tr\big(\rho(0)e^{-z^{*}\hat{a}_{1}(t)}e^{z\hat{a}_{1}^{\dag}(t)}e^{-\eta^{*}\hat{a}_{2}(t)}
e^{\eta\hat{a}_{2}^{\dag}(t)}\big).
\end{equation}
Employing the identity
\begin{equation}\label{640}
e^{\hat{A}}e^{\hat{B}}=e^{\hat{A}+\hat{B}+{1\over 2}[\hat{A},\hat{B}]}
\end{equation}
and taking into account the fact that $\hat{a}_{1}$ and $\hat{a}_{2}$ are Gaussian variables with zero mean, Eq.(\ref{639}) can be put in the form
\begin{equation}\label{641}
\phi_{a}(z,\eta,t)=exp[{-{1\over 2}(z^{*}z+\eta^{*}\eta)}exp\big[{1\over 2}\big\langle\big(z\hat{a}_{1}^{\dag}-z^{*}\hat{a}_{1}+\eta\hat{a}_{2}^{\dag}-\eta^{*}\hat{a}_{2}\big)^{2}\big\rangle\big].
\end{equation}
It then follows that
\begin{eqnarray}\label{642}
\phi_{a}(z,\eta,t)\hspace*{-2.5mm}&=exp[-{1\over 2}(z^{*}z+\eta^{*}\eta)]exp\big[\big\langle-{1\over 2}z^{*}z(\hat{a}_{1}^{\dag}\hat{a}_{1}+\hat{a}_{1}\hat{a}_{1}^{\dag})+{1\over 2}z^{* 2}\hat{a}_{1}^{2}\nonumber\\
&+{1\over 2}z^{2}\hat{a}_{1}^{\dag 2}
-{1\over 2}\eta^{*}\eta(\hat{a}_{2}^{\dag}\hat{a}_{2}+\hat{a}_{2}\hat{a}_{2}^{\dag})+{1\over 2}\eta^{* 2}\hat{a}_{2}^{2}
+{1\over 2}\eta^{2}\hat{a}_{2}^{\dag 2}\nonumber\\
\hspace*{-10mm}&+z^{*}\eta^{*}\hat{a}_{1}\hat{a}_{2}+z\eta\hat{a}_{1}^{\dag}\hat{a}_{2}^{\dag}
-z\eta^{*}\hat{a}_{1}^{\dag}\hat{a}_{2}-z^{*}\eta\hat{a}_{1}\hat{a}_{2}^{\dag}\big\rangle\big].
\end{eqnarray}
Now on account of (\ref{632}), (\ref{633}), (\ref{634}), and (\ref{635}), this equation goes over into
\begin{equation}\label{643}
\phi_{a}(z,\eta)=exp\left[-a(z^{*}z+\eta^{*}\eta)-b(z\eta+z^{*}\eta^{*})\right],
\end{equation}
where
\begin{equation}\label{644}
a={\kappa^{2}-2\varepsilon^{2}\over{\kappa^{2}-4\varepsilon^{2}}}
\end{equation}
and
\begin{equation}\label{645}
b={\kappa\varepsilon\over{\kappa^{2}-4\varepsilon^{2}}}.
\end{equation}
Finally, upon introducing (\ref{643}) into (\ref{638})
and carrying out the integration, the Q function
for the two-mode subharmonic light is found to be
\begin{equation}\label{646}
Q(\alpha_{1}, \alpha_{2})={1\over\pi^{2}}\left[u^{2}-v^{2}\right]exp\left[-u(
\alpha_{1}^{*}\alpha_{1}+\alpha_{2}^{*}\alpha_{2})-v(\alpha_{1}\alpha_{2}+\alpha_{1}^{*}\alpha_{2}^{*})\right],
\end{equation}
in which
\begin{equation}\label{647}
u=\frac{a}{a^{2}-b^{2}},
\end{equation}
\begin{equation}\label{648}
v=\frac{b}{a^{2}-b^{2}}.
\end{equation}

\section{Photon statistics}
In this section we wish to calculate the mean photon number, the variance of the photon number, and the photon number distribution for the signal-signal modes as well as the signal-idler modes.
\subsection{The mean and variance of the photon number}
We define the mean photon number of the two-mode subharmonic light by $\overline{n}=\langle\hat{a}^{\dag}\hat{a}\rangle$.
Then using Eq. (\ref{636}) and taking into account Eq. (\ref{635}), we easily find
\begin{equation}\label{649}
\overline{n}=\langle\hat{a}_{1}^{\dag}\hat{a}_{1}\rangle+\langle\hat{a}_{2}^{\dag}\hat{a}_{2}\rangle,
\end{equation}
so that in view of (\ref{632}) and (\ref{633}), there follows
\begin{equation}\label{650}
\overline{n}={4\varepsilon^{2}\over{\kappa^{2}-4\varepsilon^{2}}}.
\end{equation}
This represents the mean photon number of the signal-signal or the signal-idler modes.
We observe that the mean photon number obtained in the Appendix is half of the result given by Eq. (\ref{650}).

On account of Eq. (\ref{637}), the photon-number variance of the two-mode subharmonic light, defined by
\begin{equation}\label{651}
(\Delta n)^{2}=\langle(\hat{a}^{\dag}\hat{a})^{2}\rangle-\overline{n}^{2},
\end{equation}
can be put in the form
\begin{equation}\label{652}
(\Delta n)^{2}=\langle\hat{a}^{\dag 2}\hat{a}^{2}\rangle+2\overline{n}-\overline{n}^{2}.
\end{equation}
Now applying the fact that $\hat{a}$ is a Gaussian variable with zero mean, we get
\begin{equation}\label{653}
(\Delta n)^{2}=2\overline{n}+\overline{n}^{2}+\langle\hat{a}^{\dag 2}\rangle\langle\hat{a}^{2}\rangle
\end{equation}
and on taking into account (\ref{636}) along with (\ref{635}), we arrive at
\begin{equation}\label{654}
(\Delta n)^{2}=2\overline{n}+\overline{n}^{2}
+4\langle\hat{a}_{1}^{\dag}\hat{a}_{2}^{\dag}\rangle\langle\hat{a}_{1}\hat{a}_{2}\rangle.
\end{equation}
Hence in view of Eqs. (\ref{650}) and (\ref{634}), the photon-number variance of the two-mode subharmonic light
takes the form
\begin{equation}\label{655}
(\Delta n)^{2}={8\varepsilon^{2}\over{\kappa^{2}-4\varepsilon^{2}}}+{16\varepsilon^{4}\over{(\kappa^{2}-4\varepsilon^{2})^{2}}}
+{4\kappa^{2}\varepsilon^{2}\over{(\kappa^{2}-4\varepsilon^{2})^{2}}}.
\end{equation}

In addition, we note that the equation of evolution of the mean photon number for the pump mode can be written as
\begin{equation}\label{656}
{d\over dt}\langle\hat{b}^{\dag}\hat{b}\rangle=-i\langle[\hat{b}^{\dag}\hat{b},\hat{H}]\rangle+{1\over 2}\kappa Tr[(2\hat{b}\hat{\rho}\hat{b}^{\dag}-\hat{b}^{\dag}\hat{b}\hat{\rho}-\hat{\rho}\hat{b}^{\dag}\hat{b})\hat{b}^{\dag}\hat{b}].
\end{equation}
Then using Eq. (\ref{61}) and the fact that
\begin{equation}\label{657}
{1\over 2}\kappa Tr[(2\hat{b}\hat{\rho}\hat{b}^{\dag}
-\hat{b}^{\dag}\hat{b}\hat{\rho}-\hat{\rho}\hat{b}^{\dag}\hat{b})\hat{b}^{\dag}\hat{b}]=-\kappa\langle\hat{b}^{\dag}\hat{b}\rangle,
\end{equation}
we readily get
\begin{equation}\label{658}
{d\over dt}\langle\hat{b}^{\dag}\hat{b}\rangle=-\kappa\langle\hat{b}^{\dag}\hat{b}\rangle+\mu\hat{b}+\mu\hat{b}^{\dag}+\lambda\langle\hat{b}^{\dag}\hat{a}_{1}\hat{a}_{2}\rangle
+\lambda\langle\hat{b}\hat{a}_{1}^{\dag}\hat{a}_{2}^{\dag}\rangle.
\end{equation}
The steady-state solution of this equation is
\begin{equation}\label{659} \langle\hat{b}^{\dag}\hat{b}\rangle={\mu\over\kappa}\hat{b}+{\mu\over\kappa}\hat{b}^{\dag}+{\lambda\over\kappa}\langle\hat{b}^{\dag}\hat{a}_{1}\hat{a}_{2}\rangle
+{\lambda\over\kappa}\langle\hat{b}\hat{a}_{1}^{\dag}\hat{a}_{2}^{\dag}\rangle,
\end{equation}
so that in view of Eqs. (\ref{624}) and (\ref{634}), the mean photon number of the pump mode takes the form
\begin{equation}\label{660}
\langle\hat{b}^{\dag}\hat{b}\rangle={4\mu^{2}\over\kappa^{2}}-{2\varepsilon^{2}\over{\kappa^{2}-4\varepsilon^{2}}}.
\end{equation}
The first term represents the mean photon number of the pump mode in the absence of the subharmonic generation and the second one represents the mean photon number of light mode $a_{1}$ or light mode $a_{2}$. This is exactly what we would expect the mean photon number of the pump mode to be.

\subsection{The photon number distribution}
We finally proceed to calculate the photon number distribution for the signal-signal or the signal-idler modes.
The probability for observing $m+n$ signal photons or the probability for observing $m$ signal photons and $n$ idler photons is expressible in the form
  \begin{eqnarray}\label{661}
  &P(m,n)
  ={\pi^{2}\over m!n!}{\partial^{2m}\over\partial\alpha_{1}^{m}\partial
  \alpha_{1}^{* m}}{\partial^{2n}\over\partial\alpha_{2}^{n}\partial\alpha_{2}^{* n}}[Q(\alpha_{1},
  \alpha_{2})e^{\alpha_{1}^{*}\alpha_{1}+\alpha_{2}^{*}\alpha_{2}}]|_{\alpha_{1}=\alpha_{1}^{*}=\alpha_{2}=\alpha_{2}^{*}=0}.
  \end{eqnarray}
Thus with the aid of  Eqs. (\ref{646}) and (\ref{661}), one can write
\begin{eqnarray}\label{662}
P(m,n)\hspace*{-3mm}&=\hspace*{-3mm}&{\left[u^{2}-v^{2}\right]\over m!n!}{\partial^{2m}\over\partial
\alpha_{1}^{m}\partial\alpha_{1}^{* m}}{\partial^{2n}\over\partial\alpha_{2}^{n}\partial
\alpha_{2}^{* n}}\nonumber\\&&
\times\exp\left [(1-u)(\alpha_{1}^{*}\alpha_{1}+\alpha_{2}^{*}\alpha_{2})-v(\alpha_{1}\alpha_{2}
+\alpha_{1}^{*}\alpha_{2}^{*})\right ]_{\alpha_{1}=\alpha_{1}^{*}=\alpha_{2}=\alpha_{2}^{*}=0},\end{eqnarray}
so that on expanding the exponential functions in power series, we have
\begin{eqnarray}\label{663}
P(m,n)\hspace*{-3mm}&=\hspace*{-3mm}&{\left[u^{2}-v^{2}\right]\over m!n!}\sum_{ijk\ell}{(-1)^{k+\ell}(1-u)^{i+j}
v^{k+\ell}\over i!j!k!\ell!}{\partial^{2m}\over\partial\alpha_{1}^{m}\partial\alpha_{1}^{* m}}
{\partial^{2n}\over\partial\alpha_{2}^{n}\partial\alpha_{2}^{* n}}\nonumber\\&&
\times\left[\alpha_{1}^{i+k}\alpha_{1}^{*i+l}\alpha_{2}^{j+k}\alpha_{2}^{*j+l}\right]_{\alpha_{1}
=\alpha_{1}^{*}=\alpha_{2}=\alpha_{2}^{*}=0}.\end{eqnarray}
Now upon performing the differentiation and applying the condition $\alpha_{1}
=\alpha_{1}^{*}=\alpha_{2}=\alpha_{2}^{*}=0$, one gets
\begin{eqnarray}\label{664}
P(m,n)\hspace*{-3mm}&=\hspace*{-3mm}&{\left[u^{2}-v^{2}\right]\over m!n!}\sum_{ijk\ell}{(-1)^{k+\ell}(1-u)^{i+j}
v^{k+\ell}\over i!j!k!\ell!}\nonumber\\&&
\times{(i+k)!\over(i+k-m)!}{(i+\ell)!\over(i+\ell-m)!}{(j+k)!\over(j+k-n)!}{(j+\ell)!\over(j+\ell-n)!}\nonumber\\&&
\times\delta_{i+k,m}\delta_{i+\ell,m}\delta_{j+k,n}\delta_{j+\ell,n}.\end{eqnarray}
We note that $k=\ell=m-i=n-j$. Therefore, for $m=n$ the photon number distribution
takes the form
\begin{equation}\label{665}
P(n,n)=\left[u^{2}-v^{2}\right]\sum_{j=0}^{n}{n!^{2}(1-u)^{2j}v^{2(n-j)}
\over j!^{2}[(n-j)!]^{2}}.
\end{equation}
We observe that Eq. (\ref{665}) represents the probability to observe $2n$ signal photons or $n$ signal photons and $n$ idler photons.
\section{Quadrature squeezing}
In this section we seek to obtain the global and local quadrature squeezing for the two-mode subharmonic light.
\subsection{Global quadrature squeezing}
Here we wish to determine the global quadrature squeezing (in the entire frequency interval) for the two-mode subharmonic light. We recall that the variance of the plus and minus quadrature operators for a two-mode light is given by
\begin{equation}\label{666}
(\Delta a_{\pm})^{2}=\langle\hat{a}_{\pm},\hat{a}_{\pm}\rangle,
\end{equation}
where
\begin{equation}\label{667}
\hat{a}_{+}=\hat{a}^{\dag}+\hat{a},
\end{equation}
\begin{equation}\label{668}
\hat{a}_{-}=i(\hat{a}^{\dag}-\hat{a}),
\end{equation}
and $\hat{a}$ is the annihilation operator for the two-mode light.
Now on account of (\ref{631}) along with (\ref{636}), (\ref{667}), and (\ref{668}), we see that
\begin{equation}\label{669}
\langle a_{\pm}\rangle=0.
\end{equation}
Hence in view of this, Eq. (\ref{666}) is expressible as
\begin{equation}\label{670}
(\Delta a_{\pm})^{2}=\langle\hat{a}_{\pm}^{2}\rangle.
\end{equation}
We note that
\begin{equation}\label{671}
(\Delta a_{\pm})^{2}=\langle\hat{a}^{\dag}\hat{a}\rangle+\langle\hat{a}\hat{a}^{\dag}\rangle\pm\langle\hat{a}^{2}
+\hat{a}^{\dag 2}\rangle
\end{equation}
and applying (\ref{637}) the quadrature variance can be put in the form
\begin{equation}\label{672}
(\Delta a_{\pm})^{2}=2+2\langle\hat{a}^{\dag}\hat{a}\rangle\pm\langle\hat{a}^{2}
+\hat{a}^{\dag 2}\rangle.
\end{equation}
With the aid of (\ref{636}), one can rewrite Eq. (\ref{672}) as
\begin{equation}\label{673}
(\Delta a_{\pm})^{2}=2+2\langle\hat{a}_{1}^{\dag}\hat{a}_{1}+\hat{a}_{2}^{\dag}\hat{a}_{2}\rangle\pm\langle\hat{a}_{1}^{2}
+\hat{a}_{1}^{\dag 2}+\hat{a}_{2}^{2} +\hat{a}_{2}^{\dag 2}+2\hat{a}_{1}\hat{a}_{2}+2\hat{a}_{1}^{\dag}\hat{a}_{2}^{\dag}\rangle.
\end{equation}
Therefore on account of Eqs. (\ref{632}), (\ref{633}), (\ref{634}), and (\ref{635}), the quadrature variance takes the form
\begin{equation}\label{674}
(\Delta a_{+})^{2}=2-{4\varepsilon\over{\kappa+ 2\varepsilon}}
\end{equation}
and
\begin{equation}\label{675}
(\Delta a_{-})^{2}=2+{4\varepsilon\over{\kappa-2\varepsilon}}.
\end{equation}
We immediately note that the two-mode cavity subharmonic light is in a squeezed state and the squeezing occurs in the plus quadrature. In addition, we note that for $\kappa=2\varepsilon$ the variance of the minus quadrature diverges. We then identify $\kappa=2\varepsilon$ as the threshold condition.

Upon setting $\varepsilon=0$ in Eqs. (\ref{674}) and (\ref{675}), we find
\begin{equation}\label{676}
(\Delta a_{\pm})_{v}^{2}=2.
\end{equation}
We thus see that for $\varepsilon=0$ the cavity light is in a two-mode vacuum state in which the uncertainties in the two quadratures are equal and satisfy the minimum uncertainty relation. We wish to calculate the quadrature squeezing of the two-mode cavity subharmonic light relative to the quadrature variance of the two-mode cavity vacuum state.
We therefore define the quadrature squeezing of the two-mode cavity subharmonic light by
\begin{equation}\label{677}
S={2-(\Delta a_{+})^{2}\over 2}.
\end{equation}
Now combination of (\ref{674}) and (\ref{677}) yields
\begin{equation}\label{678}
S={2\varepsilon\over{\kappa+2\varepsilon}}.
\end{equation}
This represents the global quadrature squeezing of the cavity signal-signal or signal-idler modes. We note that at steady state and at threshold there is a 50$\%$ quadrature squeezing below the vacuum-state level.

We next seek to obtain the quadrature squeezing of the two-mode output subharmonic light. To this end, we demand that the quadrature variance of the two-mode output subharmonic light must be the sum of the quadrature variance of the transmitted two-mode cavity subharmonic light and the quadrature variance of the two-mode reflected input modes. One can then write
\begin{equation}\label{679}
(\Delta a^{out}_{\pm})^{2}=\kappa(\Delta a_{\pm})^{2}+(1-\kappa)(\Delta a_{in\pm})^{2},
\end{equation}
so that on account of (\ref{674}) and (\ref{675}) and the fact that
\begin{equation}\label{680}
(\Delta a_{in\pm})^{2}=2,
\end{equation}
we easily find
\begin{equation}\label{681}
(\Delta a^{out}_{+})^{2}=2-{4\kappa\varepsilon\over{\kappa+ 2\varepsilon}}
\end{equation}
and
\begin{equation}\label{682}
(\Delta a^{out}_{-})^{2}=2+{4\kappa\varepsilon\over{\kappa+ 2\varepsilon}}.
\end{equation}
Moreover, the quadrature variance of the two-mode output vacuum state can be written as
\begin{equation}\label{683}
(\Delta a^{out}_{\pm})_{v}^{2}=\kappa(\Delta a_{\pm})_{v}^{2}+(1-\kappa)(\Delta a_{in\pm})^{2}.
\end{equation}
Hence in view of (\ref{676}) and (\ref{680}), there follows
\begin{equation}\label{684}
(\Delta a^{out}_{\pm})_{v}^{2}=2.
\end{equation}
Now we define the quadrature squeezing of the two-mode output subharmonic light by
\begin{equation}\label{685}
S^{out}={2-(\Delta a^{out}_{+})^{2}\over 2}.
\end{equation}
Then with the aid of (\ref{681}), we obtain
\begin{equation}\label{686}
S^{out}={2\kappa\varepsilon\over{\kappa+2\varepsilon}}.
\end{equation}
Finally, at threshold and for $\kappa=0.8$, we see that the quadrature squeezing of the two-mode output subharmonic light is 40$\%$ below the vacuum-state level. We thus note that the quadrature squeezing of the two-mode output  subharmonic light is less than that of the two-mode cavity subharmonic light. This must be due to the vacuum reservoir quadrature fluctuations.

\subsection{Local quadrature squeezing}
We finally seek to calculate the local quadrature squeezing (in a given frequency interval) for the
signal-signal modes. To this end, we define the spectrum of quadrature fluctuations for the cavity signal-signal modes with central frequency $\omega_{0}$ by
\begin{equation}\label{687} S_{\pm}(\omega)=
{1\over 2\pi}\int_{-\infty}^{\infty}\langle\hat{a}_{\pm}(t),
\hat{a}_{\pm}(t+\tau)\rangle_{ss}e^{i(\omega-\omega_{0})\tau}d\tau.
\end{equation}
Upon integrating both sides of (\ref{687}) over $\omega$, we easily get
\begin{equation}\label{688}
\int_{-\infty}^{\infty} S_{\pm}(\omega)d\omega=(\Delta a_{\pm})^{2},
\end{equation}
with $(\Delta a_{\pm})^{2}$ being the steady-state global quadrature variance.
On the basis of this relation, we observe that $S_{\pm}(\omega)d\omega$ is the quadrature variance of the cavity signal-signal modes in the interval between $\omega$ and $\omega+d\omega$. The quadrature variance in the interval between $\omega'=-\lambda$ and $\omega'=\lambda$ can then be expressed as
\begin{equation}\label{689}
(\Delta a_{\pm})^{2}_{\pm\lambda}=\int_{-\lambda}^{\lambda}S_{\pm}(\omega')d\omega',
\end{equation}
in which $\omega'=\omega-\omega_{0}$.

We now proceed to obtain the spectrum of quadrature fluctuations for the cavity signal-signal modes produced by the degenerate subharmonic generator. In view of (\ref{669}), one can rewrite Eq. (\ref{687}) as
\begin{equation}\label{690}
S_{\pm}(\omega)={Re\over\pi}\int_{0}^{\infty}\langle\hat{a}_{\pm}(t)
\hat{a}_{\pm}(t+\tau)\rangle_{ss}e^{i(\omega-\omega_{0})\tau}d\tau.
\end{equation}
Moreover, upon setting $\kappa_{1}=\kappa_{2}=\kappa$ and applying (\ref{624}) in Eqs. (\ref{613}) and (\ref{614}), we get
\begin{equation}\label{691}
{d\over dt}\langle\hat{a}_{1}\rangle=-{1\over 2}\kappa\langle\hat{a}_{1}\rangle
-\varepsilon\langle\hat{a}_{2}^{\dagger}\rangle
\end{equation}
and
\begin{equation}\label{692}
{d\over dt}\langle\hat{a}_{2}\rangle=-{1\over 2}\kappa\langle\hat{a}_{2}\rangle
-\varepsilon\langle\hat{a}_{1}^{\dagger}\rangle,
\end{equation}
in which $\varepsilon$ is defined by (\ref{630}).
Now addition of Eqs. (\ref{691}) and (\ref{692}) results in
\begin{equation}\label{693}
{d\over dt}\langle\hat{a}\rangle=-{1\over 2}\kappa\langle\hat{a}\rangle
-\varepsilon\langle\hat{a}^{\dagger}\rangle,
\end{equation}
where $\hat{a}$ is defined by (\ref{636}).
Hence using Eq. (\ref{693}) and its complex conjugate, one readily finds
\begin{equation}\label{694}
{d\over dt}\langle\hat{a}_{\pm}(t)\rangle=-\eta_{\pm}\langle\hat{a}_{\pm}(t)\rangle,
\end{equation}
in which $\langle\hat{a}_{+}(t)\rangle$ and $\langle\hat{a}_{-}(t)\rangle$ are given by (\ref{667}) and (\ref{668})
and
\begin{equation}\label{695}
\eta_{\pm}={\kappa\over 2}\pm\varepsilon.
\end{equation}
We note that the solution of Eq. (\ref{694}) can be written as
\begin{equation}\label{696}
\langle\hat{a}_{\pm}(t+\tau)\rangle=\langle\hat{a}_{\pm}(t)\rangle e^{-\eta_{\pm}\tau}
\end{equation}
and applying the quantum regression theorem, we have
\begin{equation}\label{697}
\langle\hat{a}_{\pm}(t)\hat{a}_{\pm}(t+\tau)\rangle=\langle\hat{a}_{\pm}^{2}(t)\rangle e^{-\eta_{\pm}\tau}.
\end{equation}

\begin{figure}[bht]
\centering
\includegraphics[height=7.5cm,keepaspectratio]{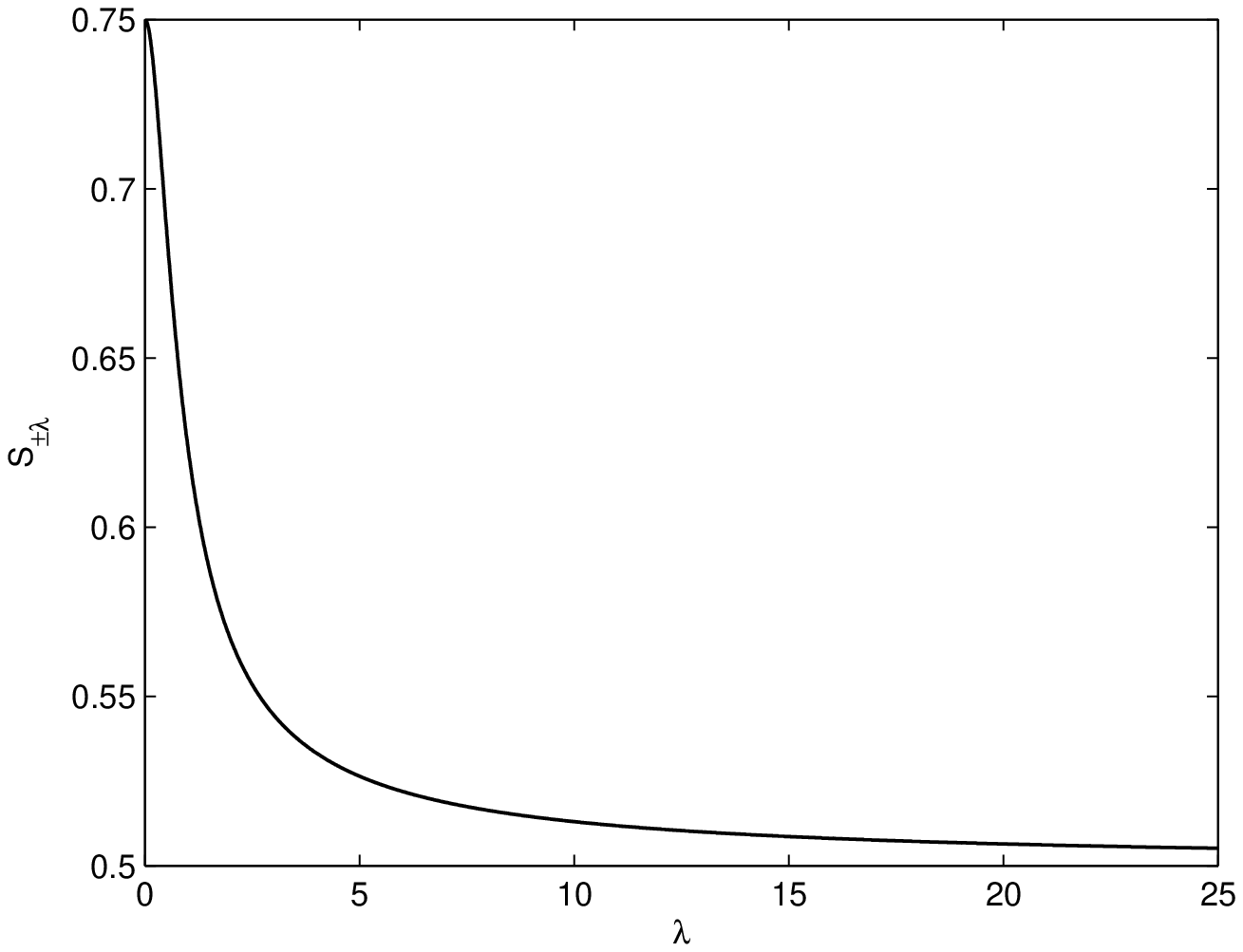}
\begin{center}
{\footnotesize{\bf Fig. 6.2}~~~~~A plot of local quadrature squeezing versus $\lambda$ [Eq. (\ref{6105})] for $\varepsilon=0.4$ and $\kappa=0.8$.}
\end{center}
\end{figure}
\noindent
Therefore, on account of (\ref{697}), one can put Eq. (\ref{690}) in the form
\begin{equation}\label{698}
S_{\pm}(\omega)={1\over\pi}(\Delta a_{\pm})^{2}
Re\int_{0}^{\infty}e^{-(\eta_{\pm}-i(\omega-\omega_{0}))\tau}d\tau,
\end{equation}
so that on carrying out the integration, the spectrum of the plus quadrature fluctuations
for the cavity signal-signal modes is found to be
\begin{equation}\label{699}
S_{+}(\omega)=(\Delta a_{+})^{2}\bigg({({\kappa\over 2}+\varepsilon)/\pi\over(\omega-\omega_{0})^{2}
+[{\kappa\over 2}+\varepsilon]^{2}}\bigg) \end{equation}
and on taking into account (\ref{674}), we get
\begin{equation}\label{6100}
S_{+}(\omega)=\bigg({({\kappa\over 2}+\varepsilon)/\pi\over(\omega-\omega_{0})^{2}
+[{\kappa\over 2}+\varepsilon]^{2}}\bigg)
\bigg({2\kappa\over{\kappa+2\varepsilon}}\bigg).
\end{equation}

Furthermore, upon integrating Eq. (\ref{6100}) in the interval between $\omega'=-\lambda$ and $\omega'=\lambda$, applying the fact that
\begin{equation}\label{6101}
\int_{-\lambda}^{\lambda}{d\omega'\over{\omega'^{2}+a^{2}}}={2\over a}tan^{-1}\bigg({\lambda\over a}\bigg),
\end{equation}
we arrive at
\begin{equation}\label{6102}
(\Delta a_{+})^{2}_{\pm\lambda}={2\over\pi} tan^{-1}\bigg({\lambda\over{\kappa\over 2}+\varepsilon}\bigg)\bigg({2\kappa\over{\kappa+2\varepsilon}}\bigg).
\end{equation}
On the other hand, on setting $\varepsilon=0$ in (\ref{6102}), we find
\begin{equation}\label{6103}
(\Delta a_{+})^{2}_{v\pm\lambda}={4\over\pi}tan^{-1}\bigg({2\lambda\over\kappa}\bigg).
\end{equation}
This represents the quadrature variance of the two-mode cavity vacuum state in the same frequency interval.
We define the quadrature squeezing of the cavity signal-signal modes in the interval between $\omega'=-\lambda$ and $\omega'=\lambda$ by
\begin{equation}\label{6104}
S_{\pm\lambda}={(\Delta a_{+})^{2}_{v\pm\lambda}-(\Delta a_{+})^{2}_{\pm\lambda}\over(\Delta a_{+})^{2}_{v\pm\lambda}}.
\end{equation}
Therefore on account of (\ref{6102}) and (\ref{6103}), we see that
\begin{equation}\label{6105}
S_{\pm\lambda}=1-{tan^{-1}\big({\lambda\over{\kappa\over 2}+\varepsilon}\big)\over 2 tan^{-1}\big({2\lambda\over\kappa}\big)}\bigg({2\kappa\over{\kappa+2\varepsilon}}\bigg).
 \end{equation}
The plot in Fig. 6.2 shows that the maximum local quadrature squeezing is 74.9$\%$ below the vacuum-state level. This occurs in the frequency interval $\lambda_{\pm}=0.05$. In addition, we note that the local quadrature squeezing approaches the global quadrature squeezing as $\lambda$ increases.

\section{Conclusion}

The mean photon number of the twin light beams with the same frequency, obtained applying the conventional Hamiltonian, is found to be half that of the twin light beams with different frequencies. This unexpected result must be due to representation of the twin light beams with the same frequency by second-order annihilation and creation operators in the conventional Hamiltonian. We then realize that the twin light beams with the same or different frequencies must be represented in the Hamiltonian by first-order annihilation and creation operators. We thus describe  the process of subharmonic generation leading to the creation of twin light beams with the same or different frequencies by a  Hamiltonian of the form indicated in Eq. (\ref{61}).

 Evidently, the mean photon number calculated using this Hamiltonian turns out to be the same for both the signal-signal and the signal-idler modes. In addition, the mean photon number of the pump mode is found to be the mean photon number of the pump mode in the absence of the subharmonic generation minus the mean photon number of either the signal-signal or the signal-idler modes. This is exactly what we would expect the mean photon number of the pump mode to be.
On the other hand, we  have established that the  maximum global quadrature squeezing of the signal-signal or the signal-idler modes is 50$\%$ below the vacuum-state level and the maximum local quadrature squeezing of the signal-signal modes is 74.9$\%$ below the same level.

\section*{Appendix: The Conventional Hamiltonian}
\setcounter{equation}{0}
Here we wish to calculate, employing the conventional Hamiltonian, the mean photon number of the twin light beams with the same frequency produced by a degenerate subharmonic generator. The process of subharmonic generation leading to the creation of the twin light beams is usually described by the Hamiltonian
\begin{align}
\hat H=i\mu(\hat b^{\dag}-\hat b)
   +{i\lambda\over 2}(\hat b^{\dag}\hat a^{2}-\hat b\hat a^{\dag 2}),\tag{A1}
\end{align}
in which $\hat{a}$ is the annihilation operator for the twin light beams, $\hat{b}$ is the annihilation operator for pump mode, $\lambda$ is the coupling constant, and $\mu$ is proportional to the amplitude of the coherent light deriving the pump mode. We may refer to a Hamiltonian of the form described by ({A1}) as second-order Hamiltonian.
We consider here the case in which the twin light beams and the pump mode are in a cavity coupled to a vacuum reservoir via a single port-mirror.
The equations of evolution of $\langle\hat{a}^{\dag}\hat{a}\rangle$ and $\langle\hat{a}^{2}\rangle$ can be written as
\begin{align}
{d\over dt}\langle\hat{a}^{\dag}\hat{a}\rangle=-i\langle[\hat{a}^{\dag}\hat{a},\hat{H}]\rangle+{1\over 2}\kappa Tr[(2\hat{a}\hat{\rho}\hat{a}^{\dag}-\hat{a}^{\dag}\hat{a}\hat{\rho}-\hat{\rho}\hat{a}^{\dag}\hat{a}
)\hat{a}^{\dag}\hat{a}]\tag{A2}
\end{align}
and
\begin{align}
{d\over dt}\langle\hat{a}^{2}\rangle=-i\langle[\hat{a}^{2},\hat{H}]\rangle+{1\over 2}\kappa Tr[(2\hat{a}\hat{\rho}\hat{a}^{\dag}-\hat{a}^{\dag}\hat{a}\hat{\rho}-\hat{\rho}\hat{a}^{\dag}\hat{a})\hat{a}^{2}], \tag{A3}
\end{align}
in which $\kappa$ is the cavity damping constant.
Now taking into account Eq. ({A1})
 and the fact that
\begin{align}
{1\over 2}\kappa Tr[(2\hat{a}\hat{\rho}\hat{a}^{\dag}
-\hat{a}^{\dag}\hat{a}\hat{\rho}-\hat{\rho}\hat{a}^{\dag}\hat{a}
)\hat{a}^{\dag}\hat{a}]=-\kappa\langle\hat{a}^{\dag}\hat{a}\rangle \tag{A4}
\end{align}
and
\begin{align}
Tr[(2\hat{a}\hat{\rho}\hat{a}^{\dag}-\hat{a}^{\dag}\hat{a}\hat{\rho}-\hat{\rho}\hat{a}^{\dag}\hat{a})\hat{a}^{2}]
=-\kappa\langle\hat{a}^{2}\rangle, \tag{A5}
\end{align}
we easily get
\begin{align}
{d\over dt}\langle\hat{a}^{\dag}\hat{a}\rangle=-\kappa\langle\hat{a}^{\dag}\hat{a}\rangle
-\lambda\langle\hat{b}^{\dag}\hat{a}^{2}\rangle-\lambda\langle\hat{b}\hat{a}^{\dag 2}\rangle \tag{A6}
\end{align}
and
\begin{align}
{d\over dt}\langle\hat{a}^{2}\rangle=-\kappa\langle\hat{a}^{2}\rangle-2\lambda\langle\hat{b}\hat{a}^{\dag}\hat{a}\rangle
-\lambda\langle\hat{b}\rangle. \tag{A7}
\end{align}

The steady-state solutions of these equations are
\begin{align}
\langle\hat{a}^{\dagger}\hat{a}\rangle=
-{\lambda\over\kappa}\langle\hat{b}^{\dagger}\hat{a}^{2}\rangle-{\lambda\over\kappa}\langle\hat{b}
\hat{a}^{\dagger{2}}\rangle\tag{A8}
\end{align}
and
\begin{align}
\langle\hat{a}^{2}\rangle=-{2\lambda\over\kappa}\langle\hat{b}\hat{a}^{\dag}\hat{a}\rangle
-{\lambda\over\kappa}\langle\hat{b}\rangle. \tag{A9}
\end{align}
Now substitution of ({24}) into Eqs. ({A8}) and ({A9}) yields
\begin{align}
\langle\hat{a}^{\dagger}\hat{a}\rangle=
-{\varepsilon\over\kappa}\langle\hat{a}^{2}\rangle-{\varepsilon\over\kappa}\langle
\hat{a}^{\dagger{2}}\rangle \tag{A10}
\end{align}
and
\begin{align}
\langle\hat{a}^{2}\rangle=-{2\varepsilon\over\kappa}\langle\hat{a}^{\dag}\hat{a}\rangle
-{\varepsilon\over\kappa}, \tag{A11}
\end{align}
where $\varepsilon$ is defined by Eq. ({30}).
Finally, on account of Eqs. ({A10}) and ({A11}), we find the mean photon number of the twin light beams to be
\begin{align}
\langle\hat{a}^{\dagger}\hat{a}\rangle={2\varepsilon^{2}\over\kappa^{2}-4\varepsilon^{2}}. \tag{A12}
\end{align}

\vspace*{3mm}
\noindent
{\bf References}

\vspace*{2mm}
\noindent
[1] J. Anwar and M.S. Zubairy, Phys. Rev. A 45, 1804 (1992).\newline
[2] G.S. Agrawal and G. Adam, Phys. Rev. A 39, 6259 (1989).\newline
[3] M.J. Collet and C.W. Gardiner, Phys. Rev. A 30, 1386 (1984).\newline
[4] G.J. Milburn and D.F. Walls, Phys. Rev. A 27, 392 (1983).\newline
[5] Fesseha Kassahun, Refined Quantum Analysis of Light (CreateSpace Independent Publishing\newline
    \hspace*{5.5mm}Platform, 2014).

\end{document}